\documentstyle[aps,amssymb,12pt]{revtex}

\begin{document}
\author{\thinspace \thinspace \thinspace \thinspace \thinspace \thinspace \thinspace
\thinspace \thinspace \thinspace \thinspace \thinspace \thinspace \thinspace
\thinspace \thinspace \thinspace \thinspace \thinspace \thinspace \thinspace
G. H. Rawitscher$^{1}$, I. Koltracht$^{2}$ and R. A. Gonzales$^{2}$}
\author{$^{1}$Physics and $^{2}$Mathematics Departments, University of Connecticut,
Storrs, CT 06268, USA}
\title{Solution of Integral Equations by a Chebyshev Expansion Method.}
\maketitle

\begin{abstract}
A new spectral type method for solving the one dimensional
quantum-mechanical Lippmann-Schwinger integral equation in configuration
space is described. The radial interval is divided into partitions, not
necessarily of equal length. Two independent local solutions of the integral
equation are obtained in each interval via Clenshaw-Curtis quadrature in
terms of Chebyshev Polynomials. The local solutions are then combined into a
global solution by solving a matrix equation for the coefficients. This
matrix is sparse and the equation is easily soluble. The method shows
excellent numerical stability, as is demonstrated by several numerical
examples.
\end{abstract}

\section{Introduction.}

In conjunction with Professor I. Koltracht and several students in the
Mathematics Department of the University of Connecticut, we are in the
process of developing a method\cite{KOL1} for solving the one-dimensional
Lippmann-Schwinger integral equations in configuration space, associated
with the corresponding Schr\H odinger equation. At a later stage we hope to
generalize the method to other types of integral equations, and attempt to
increase the dimension of the number of variables. Our method is excellently
suited for cases where either the potential has a long range with many
oscillations occurring in the wave function, or for large systems of coupled
equations in which some of the channel energies are positive and others
negative. In both situations the conventional finite difference methods
(Numerov, for example) experience difficulties because of the larger
accumulation of roundoff errors, and because of linear independence problems
in implementing the asymptotic boundary conditions in the coupled channel
case. A good physical application of our method is very likely the process
of photoassociation in the collision of cold atoms, because of the very long
range of the interaction between the atoms, and also because of the many
excited configurations which participate and which lead to many coupled
channels. Photoassociation is a relatively new field of study. It was
recently reviewed by Julienne\cite{JULIENNE}, and is observed experimentally
at various laboratories, including the University of Connecticut\cite{SG}.
The calculation of three-body reactions in coordinate space might furnish
another example because of the large distances involved, especially if we
are able to generalize our method to two dimensions, but this conjecture has
not yet been examined.

The Chebyshev expansion method is ''spectral'', i.e., the accuracy increases
faster than any inverse power of the number of mesh points employed. Our
method is based on that of Greengard and Rokhlin\cite{GR}, but differs\cite
{KOL1} from it substantially in the manner in which the local solutions in
each partition are combined into the global solution. We prefer to use
Chebyshev polynomials for the expansion basis in each partition because
these functions have excellent integral properties\cite{CC}, and the node
points can be obtained in terms of simple algebraic cosine expressions,
rather than in terms of non-analytic expressions required for the zeros of
Legendre Polynomials, for example.

\section{The Algorithm.}

For the one-channel case our integral equation method (IEM) is as follows.
The Schr\H{o}dinger equation for the partial wave function $\psi (R)$%
\begin{equation}
\left( -\frac{d^{2}}{R^{2}}+V_{L}(R)-k^{2}\right) \psi (R)=0
\end{equation}
is first transformed into an equivalent integral equation 
\begin{equation}
\psi (R)=F(R)+\int_{0}^{T}{\cal G}_{0}(R,R^{\prime })V_{L}(R^{\prime })\psi
(R^{\prime })\,dR^{\prime }.
\end{equation}
In configuration space the Green's function ${\cal G}_{0}$ is
semi-separable, i.e., it is given by the product of two independent
solutions $F_{0}$ and $G_{0}$ of the Schr\H{o}dinger equation, i.e., ${\cal G%
}_{0}(R,R^{\prime })\varpropto F_{0}(R_{<})\times G_{0}(R_{<})$. Here $R_{<}$
$(R_{>})$ is the smaller (larger) of $R$ and $R^{\prime }$, $F_{0}=\sin
(kr), $ $G_{0}=\cos (kr),$ and $F(R)$ is the driving function which is equal
to $F_{0}$ in the uncoupled case, and the upper limit in the integral is $%
T=R_{Max}$. Physicists would be tempted to use instead of $F_{0}$ and $G_{0}$
the L-dependent Riccati-Bessel functions $F_{L}$ and $G_{L}$, but because
the spectral expansion technique is very robust, it allows us to place the $%
L(L+1)/R^{2}$ term into the potential $V_{L}$%
\begin{equation}
V_{L}(R)=L(L+1)/R^{2}+\bar{V}(R),
\end{equation}
without loss of accuracy\cite{KOL1}.

We avoid the occurrence of large non-sparse matrices by: a) Dividing the
integration interval $[0,R_{Max}]$ into $m$ partitions $%
[0,b_{1}],[b_{1},b_{2}],...\;[b_{m-1},R_{Max}]$. The size of each partition
can be arbitrary, but two or three partitions per local wave number is
optimum. b) Solving the integral equation separately in each partition $i$
for two functions $Y$ and $Z$ 
\begin{eqnarray}
Y_{i}-\int_{b_{i-1}}^{b_{i}}{\cal G}_{0}(R,R^{\prime })V_{L}(R^{\prime
})Y_{i}(R^{\prime })dR^{\prime } &=&F(R),\;\;\;b_{i-1}\leq R\leq b_{i} \\
Z_{i}-\int_{b_{i-1}}^{b_{i}}{\cal G}_{0}(R,R^{\prime })V_{L}(R^{\prime
})Z_{i}(R^{\prime })dR^{\prime } &=&G(R),\;\;\;b_{i-1}\leq R\leq b_{i}.
\end{eqnarray}
The method of solution for this step leads to small non-sparse matrices. It
consists in expanding the unknown solutions into Chebyshev polynomials and
making use of their convenient properties [4-6]. c) ''Stitching'' together
the separate solutions into the global one by means of the expressions,
valid in each partition 
\begin{equation}
\psi (R)=A_{i}Y_{i}(R)+B_{i}Z_{i}(R),\;\;\;b_{i-1}\leq R\leq b_{i}.
\end{equation}
This ''stitching'' procedure leads to a large but sparse matrix. It produces
a seamless smooth continuation of the function $\psi $ from one partition
into the next, and is a consequence of the semi-separability of ${\cal G}$.
As a consequence of this property the coefficients $A$ and $B$ obey\cite
{KOL1} the matrix equation 
\begin{equation}
\left[ 
\begin{array}{cccccc}
I & M_{12} &  & 0 & 0 & 0 \\ 
M_{21} & I & M_{23} & 0 & 0 & 0 \\ 
& M_{32} & \ddots &  &  &  \\ 
0 & 0 &  & \ddots &  &  \\ 
0 & 0 &  &  & I & M_{m-1,m} \\ 
0 & 0 &  &  & M_{m,m-1} & I
\end{array}
\right] \left( 
\begin{array}{c}
\alpha _{1} \\ 
\alpha _{2} \\ 
\vdots \\ 
\vdots \\ 
\alpha _{m-1} \\ 
\alpha _{m}
\end{array}
\right) =\left( 
\begin{array}{c}
0 \\ 
0 \\ 
\vdots \\ 
\vdots \\ 
0 \\ 
u
\end{array}
\right) .
\end{equation}
where the $\alpha _{j}$ and $u$ are two-rowed columns 
\begin{equation}
\alpha _{j}=\left( 
\begin{array}{c}
A_{j} \\ 
B_{j}
\end{array}
\right) ;\;\;\;\;\;\;\;\;\;u=\left( 
\begin{array}{c}
1 \\ 
0
\end{array}
\right) ;\;\;\;\;\;\;\;\;\;0=\left( 
\begin{array}{c}
0 \\ 
0
\end{array}
\right) .
\end{equation}

For the one-channel case the $M_{i,j}$ are $2\times 2$ matrices whose
elements are formed out of overlap integrals between combinations of the
functions $Y,Z,$ the potential $V$ and the functions $F$ and $G$ \cite{KOL1}%
. We call the matrix formed out of the blocks of the $M_{ij}^{\prime }$s the
''Big Matrix''. It is sparse, and of dimension $2m\times 2m,$ where $m$ is
the number of partitions, while the matrices required to solve the equations
for the $Y$ and $Z$ in each partition are of the dimension $n\times n,$
where $n$ is the number of Chebyshev points in each partition. Following
standard practice, that number is chosen to have the value $n=16.$ The end
result of the whole calculation is the set of coefficients of the Chebyshev
polynomials for the functions $Y$ and $Z$ in each partition, as well as the
coefficients $A$ and $B$ for each partition. The calculations in one
partition of the $Y,Z,$ and the overlap integrals required for the
construction of the $M_{ij}$ can be performed independently of each other.
Hence this part of the calculation can be carried out in a parallel computer
architecture. Because of the simple structure of the ''Big Matrix'', the
solution of Eq. (7) can be performed by simple pivoting, and does not
require special big matrix techniques. The values of $\psi $ and its
derivative $d\psi /dR$ can be obtained at any arbitrary point from the
solution given by Eq. (6), because the values and the derivatives of
Chebyshev functions are known analytically. At the end point $R_{Max}=T$ the
expression for $d\psi /dR$ is especially simple.

When the Schr\H odinger Eq. is augmented to a system of coupled equations
with $N$ channels, then each $F_0$ and the corresponding $Y$ is augmented
into $N$ column vectors of length $N,$ and similarly for the $G^{\prime }s$
and $Z^{\prime }s.$ The size of the block matrices $M_{ij}$ increases
correspondingly to the dimension $2N\times 2N$, but the sparse structure of
the ''Big Matrix'' remains the same. For the channels which are closed, the
functions $F_0$ and $G_0$ are replaced by $\sinh (\kappa R)$ and $\exp
(-\kappa R),$ respectively, where $\kappa $ is the imaginary part of the
asymptotic wave number in the closed channel. Numerical experiments \cite
{KOL2} for $L=0$ show that the stability and high precision of the results
is maintained in this case also, provided that a scaling procedure is
introduced which reduces the numerical disparity between the two
exponentials $\exp (\pm \kappa R)$ for large values of $\kappa R.$ If in the
positive energy channels the angular momentum $L\neq 0,$ it is still
advantageous to use for the channel Green functions the undistorted ${\cal G}%
_0$ ones, even though one now has to solve the coupled integral equations as
many times as there are open channels, each time with a different driving
function $F.$ The big matrix $M$ is the same in each case, the only change
is in the vector $u$ on the right hand side of Eq. (7), and one can show\cite
{KOL2} that the corresponding solutions are linearly independent
asymptotically. In the next section several examples illustrate the
numerical properties of our method.

\section{Accumulation of Roundoff Errors.}

This property is demonstrated by a numerical example for an uncoupled
channel case with $L\neq 0,$ but in which the only potential present is $%
L(L+1)/R^{2}$. In this case the solution is a Riccati-Bessel function $%
F_{L}(R)=kRj_{L}(kR).$ The numerical solution $\psi $ of Eq. (2) is
proportional to $F_{L},$ and the constant of proportionality is obtained
from the Wronskian of $\psi $ with $G_{L}$ at $R=R_{Max}.$ The calculations
are done in double precision on a IBM 3090 Mainframe machine. The functions $%
G_{L}$ and $F_{L}$ at $R_{Max}$ are obtained from the International
Mathematical Scientific Library (IMSL). The error in the numerical solution $%
\psi $ is obtained by comparing it with the IMSL values of F$_{L}$ at many
points in the interval $[0,R_{Max}],$ and the maximum of the absolute values
of all these differences is denoted as ''Error''. This error is plotted in
Fig. 1 as a function of $N,$ the total number of integration points in $%
[0,R_{Max}].$ For the IEM that number is equal to 16 times the number of
partitions $m,$ while for the Numerov method it is $R_{Max}$ divided by the
uniform mesh size. In this example with $k=1fm^{-1},$ $R_{Max}=50fm\,\,$and $%
L=6,$ one sees that the error rapidly drops as a function of $N,$ and
reaches a minimum (equal to machine accuracy in this case) at the point
where the roundoff error overtakes the truncation error. Beyond that minimum
the rise is a measure of the rapidity of accumulation of roundoff error. One
sees from the figure that the Numerov error decreases much more slowly with $%
N,$ and the accumulation of round-off error increases much faster. This is
the reason why the best accuracy for the Numerov method is much worse. A
variable step size improved Numerov method due to Raptis and Cash\cite{RC}
accumulates roundoff error more slowly than the Numerov method, but still
does not reach the quality of the IEM.

A more taxing example with $L=8,$ $k=40fm^{-1}and$ $R_{Max}=50$ (not
illustrated here) shows a similarly slow accumulation of the roundoff error
for the IEM. In this case there are 580 nodes in the wave function, and the
smallest error is one order of magnitude larger than machine accuracy\cite
{KOL1}. The corresponding value of $N$ is $12800,$ which corresponds to 800
partitions in the $[0,R_{Max}]$ interval, or 20 Chebyshev points per half
wavelength. The corresponding number for best accuracy ($\simeq 10^{-8})$ of
the Numerov method is 640 points per half wavelength, larger by a factor of
30. There results are summarized in the Table below.

\begin{center}
Maximum Accuracy of the Riccati-Bessel Function.

\begin{tabular}{|l|l|l|l|}
\hline
&  & \# oscill. & \# of points/(local $\lambda )$ \\ \hline
k(fm$^{-1})$ & $\lambda (fm)$ & 0%
\mbox{$<$}%
R%
\mbox{$<$}%
50 & IEM \hspace{0.2in}Numerov \\ \hline
1 & 6.3 & \ \ \thinspace 8 & 50 \hspace{0.1in} \hspace{0.1in} \hspace{0.1in}%
1260 \\ \hline
40 & 0.16 & 310 & 40 \hspace{0.1in} \hspace{0.1in} \hspace{0.1in}1280 \\ 
\hline
\end{tabular}
\end{center}

\section{Implementation of Boundary Conditions.}

For the case of one channel the asymptotic boundary conditions $\psi
(R)\thickapprox F_L(R)+wG_L(R)$ are easily obtained in both the IEM and the
finite difference methods (FDM), by simple normalization at the end point.
However if there are two or more coupled channels, then the various
supposedly independent wave functions obtained in the FDM by starting the
solutions near the origin independently, can loose a large part of their
independence near the matching point, and the process of obtaining the
appropriate linear combination which satisfies the desired boundary
condition looses accuracy. This is not the case for the IEM, as will now be
demonstrated for the case $L=0.$

In this example\cite{KOL2} we have two channels, both with the same positive
energy $E=k^2fm^{-2},$ and all potentials (coupling as well as diagonal) are
of exponential form $V_0\exp (-r/a),$ with the same decay parameter $a$ and
the same magnitude of the strength $V_0.$ For the single channel $L=0$ case
there exists an analytic solution given in terms of Bessel functions $J_\nu
(y)$ of imaginary index $\nu =\pm 2iak,$ and argument $y=2a\sqrt{-V_0}\exp
(-r/2a),$ which can be generalized to the coupled channel case when the
energies in all the channels are equal. In our example $\mid V_0\mid =5/%
\sqrt{2}fm^{-2}$ and $a=4fm$. Channel 1 is the incident channel, with $%
V_0>0, $ and $\psi _1(R)\thickapprox F_0(R)+w_1G_0(R),$ while in channel 2 $%
V_0<0$ and $\psi _2(R)\thickapprox w_2G_0(R).$ The value of $R_{Max}$ is $%
180fm$ and $140fm,$ respectively, in the IEM and Numerov calculations, again
carried out in double precision on the IBM mainframe. For each value of the
wave number $k$ the number of points in each method of calculation is varied
until the combined error in $w_1$ and $w_2$ is a minimum, and this minimum
error is displayed on the vertical axis in Fig. 2. The reason why the
Numerov error (open circles) increases as $k$ decreases is because of the
lack of independence of the two solutions. This is shown either by an
analytical argument\cite{KOL2}, valid in this exponential case, or can seen
from the fact that the Numerov error for $w$ in the uncoupled channels is
much less dependent on $k.$ The latter is shown by the lines with squares or
triangles, labeled ''Attr.'' and ''Rep.'', respectively. The error in the
IEM is even more independent on $k,$ and is also much smaller since the
boundary conditions are automatically built into the Green's functions.

The case with the same exponential potentials in which however the energy in
the second channel is negative has also been examined. Since an analytical
solution does not exist in this case, the accuracy of the solution has to be
inferred from the stability of the values of $w_1$ and $w_2.$ For the IEM
the values of both $w_1$ and $w_2$ are stable to 13 sign figures, while for
the Numerov method the value of $w_2$ is correct to less than 4 significant
figures for $kR_{Max}>25,.$while $w_1$ is correct to about 7.

\section{Summary and Conclusions.}

We have demonstrated a method for solving the Lippmann-Schwinger integral
equation in configuration space which is linear in the number of the
integration points, and which is numerically very stable. The stability of
the IEM method is due to two factors: 1. The solution of integral equations
lead to a smaller accumulation of roundoff errors than the finite difference
methods of solving a differential equation, and 2), The IEM is a spectral
method, which has an inherently higher accuracy than finite difference
methods.

Furthermore the IEM is easy to implement: a) {\em The} {\em boundary
conditions} on the channel wave functions are automatically incorporated
through the Green's functions; b) {\em The size of each partition} $%
b_{i-1}-b_i$ can be assigned independently of the sizes of the other
partitions; c) The {\em overlap integrals} required for the construction of
matrix elements can be obtained with the Gauss-Legendre quadrature because
the functions in the integrand, being given in terms of Chebyshev
expansions, can be calculated easily for any given sets of points, no
interpolation being required. The Curtis-Clenshaw (C-C) quadrature\cite{CC}
is also a good option (see Table 2 in Ref.\cite{KOL1}) because the wave
functions and potentials are already known at the appropriate Chebyshev
points, and certainly exceeds the efficiency of equidistant point
methods.

\bigskip\ 

\subsection{Figures}

Fig. 1 Accumulation of roundoff\newline
errors for a Riccati-Bessel\newline
function, as a function of \newline
the number of integration steps N.\bigskip\ 

Fig. 2. Accuracy of the R-matrix\newline
elements for two channels\newline
coupled by exponential potentials,\newline
as a function of the wave number.

\end{document}